\documentstyle[11pt,paspconf,psfig,epsf]{article}
                                                 
\def \hi {H\,{\sc i~}}
\def \rosat {{\it ROSAT}$\,$}

\begin{document}
                
\title{Hydrostatics of the Galactic halo}
                                         
\author{Peter M.W. Kalberla and J\"urgen Kerp}
\affil{Radioastronomisches Institut der Universit\"at Bonn, Auf dem
H\"ugel 71, D-53121 Bonn, Germany}

\begin{abstract}

We investigated a hydrostatic equilibrium model of the Milky Way following 
Parker (1966), to constrain the large scale properties of the interstellar 
medium.
In our approach we found an excellent agreement
between our simple hydrostatic equilibrium model of the Milky Way and the
recent all-sky survey data rangeing from the $\gamma$-ray to the radio regime.
                                                          
On large scales the galactic disk-halo system is found to be stable against
Parker-instabilities.
Pressure support from the Galactic disk is essential to stabilise the halo.
In particular the diffuse ionised gas layer acts as a disk-halo interface.
                                                                               
Assuming that the distribution of the soft X-ray emitting plasma traces the
gravitational potential, we derived the dark matter content of the Milky Way
to be about $M \simeq  2.8 \; 10^{11}\,{\rm M_{\odot}}$.
Our findings are consistent with the rotation curve of the Galaxy.
                                            
\end{abstract}

                                  
\section{Introduction}
                      
Since the early
fifties it is known from optical polarisation studies, that 
magnetic fields are constituents of the Galactic interstellar medium.
The magnetic field strength is about a few $\mu$G. 
Radio continuum observations clearly indicate synchrotron radiation originating high above the
Galactic plane.
Thus, magnetic fields and cosmic rays are obviously constituents of the Galactic halo.
But what is about the gas within the Galactic halo? Already Parker (1966) showed, that magnetic fields are always associated with the gaseous phase. 
Investigations in the UV-range show that highly ionised gas is common within the halo, 
but it is a long way from a pencil beam to the whole volume of the Galactic halo.

Recent investigations of the \rosat soft X-ray
background data indicated the existence of a pervasive X-ray emitting plasma 
($T\,\simeq\,1.5 \; 10^6$\,K) with a vertical scale height of
about 4.4\,kpc (Pietz et al. 1998) within the halo. Moreover, a sensitive analysis of the
Leiden/Dwingeloo \hi survey
gave evidence for a \hi emission component with a high velocity dispersion
of 60 $\rm km\,s^{-1}$ (Kalberla et al. 1998) also detectable across the entire sky.
The discovery of both gas components within the Galactic halo encouraged us to study the
hydrostatic equilibrium model of the Milky Way once again.
For this approach we studied recent all-sky surveys of \hi gas,
soft X-ray radiation,
high energy $\gamma$-ray emission, and
radio-continuum emission.
                                           
%
\begin{figure}[h]
\centerline{
\psfig{figure=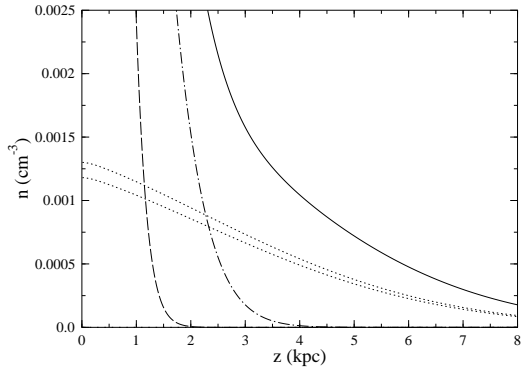,width=6.5cm,bbllx=80pt,%
bblly=150pt,bburx=460pt,bbury=430pt}}
\caption[]{
Vertical density distribution of the gas components in the solar
vicinity. The dotted lines mark both halo gas components (lower: neutral gas,
upper: ionised gas).
The neutral disk gas is indicated by the dashed line. The DIG
layer is represented by the dot-dash
line; the sum of all components is given by the solid line.
\label{fig2} }
\end{figure}

\section{A gaseous halo}
                        
To describe the large-scale properties of the Milky Way 
we used the approach of a hydrostatic halo model, as proposed by
Parker (1966). To describe the gaseous disk-halo system, we identified
3 main constituents of the Galactic interstellar medium, namely: 
the neutral interstellar gas with $h_z$ = 400 pc
(Dickey \& Lockman, 1990),
the diffuse ionised gas (DIG) with $h_z$ = 950 pc (Reynolds, 1997),
and halo gas with $h_z$ = 4.4 kpc (Kalberla et al. 1998, and Pietz et al. 1998).
The major difference to the previous studies of the hydrostatic equilibrium of the
Milky Way (e.g. Bloemen 1987, Boulares \& Cox 1990) is the detailed knowledge about the
gas phase in the Galactic halo.
In particular, the X-ray plasma in combination with the \hi high-velocity dispersion component
adds major physical parameters to our model.

Fig. 1 displays the vertical density distributions
of the gas phases (diffuse neutral and ionised gas as well as the X-ray plasma) in
the solar vicinity. Fig. 2 gives an impression on the radial density
distribution represented by the parameter $g_1$ according to Taylor \& Cordes (1993) with $A_1 = 15$ kpc.
                                                          
%
\begin{figure}[h]
\centerline{
\psfig{figure=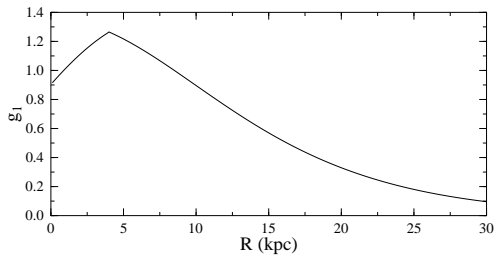,width=6.5cm,bbllx=80pt,%
bblly=150pt,bburx=460pt,bbury=330pt}}
\caption[]{
The parameter $g_1$ is a measure for the radial distribution for the gas density at $z$ = 0 kpc.
The radial scale length is about $A_1 = 15$ kpc.
\label{fig3} }
\end{figure}
\section{Gas, magnetic field and cosmic rays in equilibrium }
                                                             
Following Parker's (1966) suggestion, we studied whether gas, magnetic
fields and cosmic rays in the Galactic halo may be in pressure equilibrium.
Indeed, hydrostatic equilibrium models fit the all-sky-averaged observations best.
In detail we tested the hydrostatic equilibrium model by modelling the Galactic synchrotron
emission at 408 MHz as observed by Haslam et al. (1982), the 
$\gamma$-ray emission as observed with {\it EGRET} at energies $>$ 100 MeV
(Fichtel et al. 1994) as well as by modelling the Galactic X-ray plasma distribution deduced from
the \rosat all-sky survey data (Pietz et al. 1998). A detailed discussion of the model calculations
and a quantitative comparison with the observations are beyond the scope of this
contribution; for details we refer to Kalberla \& Kerp (1998).
Here we summarise the main features of the model.
We found a pressure equilibrium between gas,
magnetic fields and cosmic rays within the Galactic halo.
The magnetic field of the Galactic halo is globally regularly ordered and orientated
parallel to the Galactic plane.
In contrast to the halo the magnetic field within the disk is highly irregular and has
only 1/3 of the gas pressure.
                                                      
%
\begin{figure}[h]
\centerline{
\psfig{figure=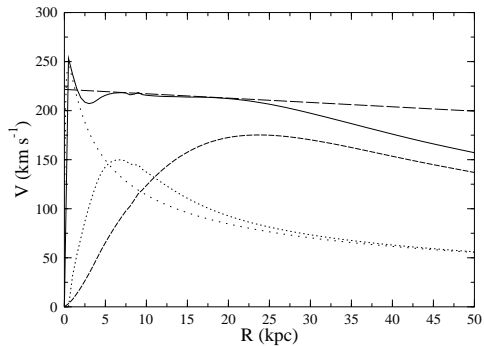,width=6.5cm,bbllx=80pt,%
bblly=150pt,bburx=460pt,bbury=430pt}}
\caption[]{
The rotation curve of the Milky Way. The dotted lines represent the stellar disk and the Galactic
bulge. The short-dashed line marks the contribution of the dark matter, quantitatively traced
by our modelling of the Galactic halo.
The sum of all three components is given by the solid line, which has to be compared with the
Galactic rotation curve deduced by Fich et al. (1990) marked by the long-dashed line.
\label{_dm_} }
\end{figure}

\section{Mass distribution and gravitational forces}
                                                    
For a galaxy in hydrostatic equilibrium the 3-D distributions of
gas pressure, density and gravitational potential are identical in
size and shape.
Accordingly, we can utilise our parameterisation of the Milky Way to
deduce the gravitational potential {\em and} the dark matter
content.
                        
%
\begin{figure}[t]
\centerline{
\psfig{figure=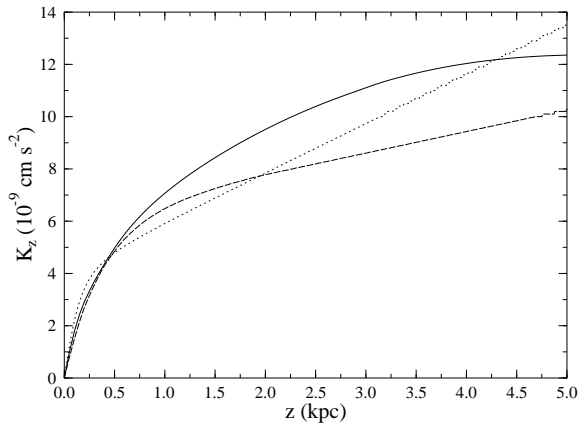,width=7.5cm,bbllx=80pt,%
bblly=150pt,bburx=460pt,bbury=430pt}}
\caption[]{
The gravitational acceleration $K_z$ in the solar neighbourhood (solid line)
deduced from the presented hydrostatic equilibrium model including the 
derived dark matter content.
For comparison $K_z$ derived by Kuijken \& Gilmore (1989) (dotted line)
and Bienam\'e et al. (1987) (dashed line) is given for $|z| < 5 $ kpc.
The  differences between the individual $K_z$ curves for $z > 1.5 $ kpc are
due to different model assumptions concerning the Galactic halo.
\label{_dm_} }
\end{figure}
                                      
In a simple view, the Galaxy consists of 3 main parts: the Galactic bulge,
the stellar disk with a radial scale length of 4.5 kpc and the gaseous halo as
described above. Assuming that the gaseous halo
traces the dark matter distribution we optimised the
density of the gaseous halo component until the rotation velocity 
of the modelled distribution was in quantitative agreement with the observed
rotation velocities (i. e. Fich el al., 1990) within galactocentric radii
3 $ < R < $ 25 kpc.
Fig.\ 3 shows the corresponding rotation curve.
The total mass of the Galaxy within $R$ = 50 kpc derived from our model is
M=$2.8 \cdot 10^{11} M_{\odot}$, consistent with
M=$ 2.4 \cdot 10^{11} M_{\odot}$ 
(Little \& Tremaine, 1987) and also within the uncertainties with
the results of Kochanek (1996) of M=$4.9 \cdot 10^{11} M_{\odot}$.

%
\begin{figure}[h]
\centerline{
\psfig{figure=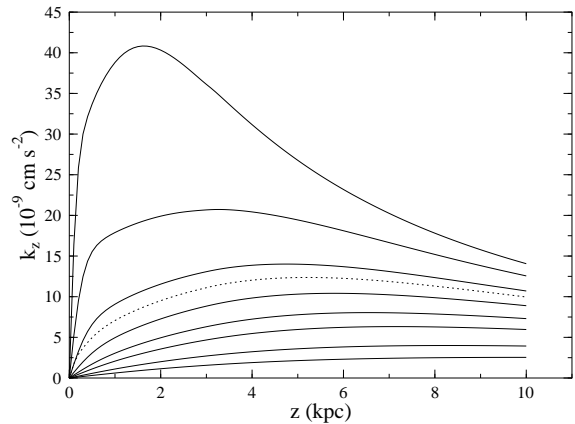,width=7.5cm,bbllx=80pt%
,bblly=150pt,bburx=460pt,bbury=430pt}}
\caption[]{
The gravitational acceleration $K_z$ according to our model plotted as a function of
galactocentric distances $R$ = 2.5, 5, 7.5, 8.5, 10, 12.5, 15, 20 and 25
kpc (from top to bottom). The solar radius is marked by the dotted line.
\label{_dm_} }
\end{figure}
                                      
In Fig. 4 we show the gravitational acceleration $K_z$ in the solar
neighbourhood as a function of $z$ deduced from our model in comparison to that of Kuijken \& Gilmore (1989)
and Bienam\'e et al. (1987). Within vertical distances of $z < 1$ kpc
our model (solid line) is in excellent agreement with
$K_z$ derived by Kuijken \& Gilmore (1989) (dotted line)
and Bienam\'e et al. (1987) (dashed line). The differences at larger $k_z$ distance is
because of different model assumptions on the dark matter distribution.
The turn-over of our model about 5 kpc above the disk is because of the radial dependence of $K_z$,
as shown in Fig.\,5 (the solar radius is marked by the dotted line).

\section{Summary and conclusion}
                                
The large scale properties of the Galactic halo are very well modelled
assuming that the main constituents of the
interstellar matter, the gas, the magnetic fields, and the cosmic rays
are in hydrostatic equilibrium.
We analysed recent all-sky surveys of \hi gas, soft X-ray radiation,
high energy $\gamma$-rays and synchrotron radiation
to test the model assumptions. In general we find
good quantitative agreement between model and data. The assumption that the gaseous halo
traces the dark matter in the Galaxy leads to a total mass which
is consistent with the observed rotation curve.
                                                          

\clearpage
Leo Blitz:
   I'm concerned about the reality of the 60 km/s component.  First,
   I don't see how it's possible to have a static ISM with such highly
   supersonic velocities.  Second, the component is not seen in the Bell
   Labs survey which has very good instrumental sidelobe response.  I
   can understand how baselines can introduce features in a spectrum,
   but I don't see how a bad baseline can coincidentally remove a
   feature from all over the sky.
                                 
Peter Kalberla:
The reality of the 60 km/s component was discussed in detail by
Kalberla et al. (1998, A\&A 332, L61). For a comparison between the Bell Labs
and Leiden/Dwingeloo surveys concerning broad emission lines see
Kalberla et al. (1997, in Proceedings of the IAU Colloquium No. 166
``The Local Bubble and Beyond'', eds. D. Breitschwerdt, M.J.
Freyberg, J. Tr\"umper, Lecture Notes in Physics 506, 475).
The fact that the observed line width of 60 km/s exceeds the thermal H\,{\sc i~}
line width significantly does not imply that the HI gas is in supersonic motion.
The H\,{\sc i~} gas in the halo has a volume filling factor of 0.12 only.
One has to consider the motion of individual H\,{\sc i~} eddies with respect to the
surrounding plasma. Since the sound velocity of this plasma is more than twice
as high as the typical H\,{\sc i~} eddy velocity of 60 km/s, the motions
of the H\,{\sc i~} clumps are clearly {\em sub}sonic.
Dissipative cloud-cloud collisions are of little relevance
in a multiphase halo.
                     
\end{document}